\documentstyle[12pt,epsfig]{article}
\newcommand\slv{v\kern-5pt\raise1pt\hbox{$\scriptstyle/$}\kern1pt}

\def\be{\begin{equation}}
\def\ee{\end{equation}}
\def\bq{\begin{eqnarray}}
\def\eq{\end{eqnarray}}  
\newcounter{saveeqn}
\newcounter{App} 

\begin{document}
\thispagestyle{empty}
\begin{flushright}
WUE-ITP-98-036\\
\end{flushright}
\vspace{0.5cm}
\begin{center}
{\Large \bf 
Top quark production near threshold:\\[.3cm] 
NNLO QCD correction }\\[.3cm]
\vspace{1.7cm}
{\sc \bf Oleg ~Yakovlev$^{a}$}\\[1cm]
\begin{center} \em  
Institut f\"ur Theoretische Physik, \\ 
Universit\"at W\"urzburg, D-97074 W\"urzburg, Germany \\
\vspace{4mm}
\vspace{4mm}
\end{center}\end{center}
\vspace{2cm}
\begin{abstract}\noindent
{ We calculate the cross section of 
the process $e^+e^- \to t\bar t$ near threshold by 
resumming Coulomb-like terms with next-to-next-to-leading   
(NNLO) accuracy. The nonrelativistic Green function
formalism and the method of ``direct matching'' are used.
The NNLO correction turns out to be large, of the same 
size as the NLO correction. It changes the position and the normalization 
of the $1S-$peak. The obtained results are compared with 
results existing in the literature. 
} 
\end{abstract}
\vspace{1cm}
\centerline{\em PACS numbers: 12.38.Bx, 12.38.Cy, 13.85.Lg, 14.65.Ha}

\vspace{1cm}
                    
\vspace*{\fill}

\noindent $^a${e-mail: iakovlev@physik.uni-wuerzburg.de}
\newpage

{\bf 1. Introduction.}\\ 

A detailed study of the process $e^+e^- \to t\bar t$ 
will be performed at the Next Linear Collider \cite{Rev}.
As a result of this study, precision measurements of
the top quark mass $m$ and width $\Gamma$ and 
of the QCD coupling constant $\alpha_s(m)$ can be done at NLC. 

The cross section of the top quark production at LO and NLO was studied 
in detail in \cite{FK,PS,JKTS,Sumino,MYa}. 
The general approach for the  calculation of the cross section 
$e^+e^- \to t\bar t$ was developed in  \cite{FK}, which pointed out that 
the large width of the top quark serves as a cutoff for 
long-distance effects in this problem. This allows for
reliable predictions for the cross section to be obtained by using 
perturbative QCD. However,
it is well known that the standard 
QCD perturbation theory does not work 
in the threshold region. 
Even at LO approximation one has to resum all 
$(\alpha_s/v)^N$ terms.
The result of such resummation is proportional 
to the nonrelativistic Coulomb Green function at the origin. 
At NLO and NNLO one has to resum all 
$(\alpha_s/v)^N(1,\alpha_s,v,\alpha_s^2,\alpha_sv,v^2$) terms
in order to calculate 
the cross section in the threshold region.

The approach to the NNLO calculations was
suggested in \cite{Hoang} and consists of 
two steps. First, one calculates the cross section in NNLO order
far from the threshold, in the region $\alpha_s \ll v \ll 1$, 
( $v=\sqrt{1-4m^2/s}$ is the relative 
velocity of top quark and antiquark),
 where the resummation of the Coulomb singularities is not needed.
Then one considers 
the nonrelativistic 
cross section using an infrared cutoff and matches it 
with the cross section in the region  $\alpha_s \ll v \ll 1$.

The Abelian part of the corrections to the 
cross section near threshold proportional to
the structures $C_F^2$ and $C_FTN_L $ were calculated in \cite{Hoang} 
and \cite{Kuehn}, respectively.
The complete NNLO correction to the cross section for the 
threshold in the  
region $\alpha_s\ll v \ll 1$ was
calculated in \cite{CM,Beneke}.
The NNLO correction to the cross section of top 
production near threshold was 
considered recently in two papers \cite{HT} and \cite{MY} by using 
two alternative approaches.  
 The difference of the numerical results between \cite{HT}
and \cite{MY} is expected to be small,
but actually the difference is large and amounts to $5-15\%$.

The main purpose of this letter is 
to calculate the cross section of the process 
$e^+e^- \to t\bar t$ in the energy region close to the threshold by
resumming Coulomb-like terms with next-to-next-to-leading   
accuracy.
 
We use two alternative approaches  
and clarify the origin of the difference 
between \cite{HT} and \cite{MY}. 
Our final numerical result differs from both 
calculations (\cite{HT} and \cite{MY}). 


In section 2, we discuss basic elements needed to 
calculate the NNLO cross section. 
 We consider the matching procedure and fix the short 
distance coefficient in section 3. 
In section 4, we present the final formula for the cross section and
results of the numerical analysis. \\

{\bf 2. Basic elements of $R_{NNLO}$.}\\

 The cross section of the process $e^+e^-\to t \bar t$ 
in the near threshold region can be obtained by using the 
optical theorem and an expansion of the vector current 
 $\bar t \gamma_{\mu} t$. The cross section reads 
\be
\label{cross}
R=\sigma(e^+e^-\to t\bar t)/\sigma_{pt}=
e^2_QN_c\frac{24\pi}{s} C(r_0){\rm Im }\Big[(1-\frac{\vec p^2}{3m^2})
G(r_0,r_0|E+i\Gamma ) \Big] |_{r_0\to 0},
\ee
where $\sigma_{pt}=4\pi\alpha^2/3s$, 
$e_Q$ is the electric charge of the top quark,
$N_c$ is the number of colors, $\sqrt{s}=2m+E $ is the total 
energy of the quark-antiquark system, $m$ is the top quark pole mass and 
$\Gamma$ is the top quark width.
$G(\vec r, \vec r'| E+i\Gamma)$ is the nonrelativistic
Green function, which satisfies to  the Schr\"odinger equation
\be
(H-E-i\Gamma)G(\vec r, \vec r'|E+i\Gamma)=\delta (\vec r - \vec r').
\ee
The Green function is divergent at the origin $r\to 0$ because  
the potential contains $1/r^2$ terms. These singularities are
regularized and 
factorized into the short distance coefficient $C(r_0)$.

In eq.(1) we have already used a complex energy,  $E+i\Gamma$, 
 in the Green function, according to \cite{FK}. 
This structure appears in the top quark propagator 
through the top quark self energy at LO, NLO and NNLO. 
The non-factorizable corrections were studied at NLO 
in \cite{MYa,Fadin,Sumino} and it was shown 
that they cancel in the total cross section. 
It was argued in \cite{FKM1} that the cancellation appears in all 
$\alpha_s$ orders; therefore we do not consider nonfactorizable 
corrections here.\\

{\bf 2a. Hamiltonian and QCD potential.}\\

The nonrelativistic Hamiltonian of the heavy quark-antiquark system reads 
\be\label{www}
 H = H_0+ W(r), \qquad
 H_0 = \frac {\vec{p}^2}{m} +V(r), 
\ee
here $V(r)$ is a static QCD potential of the heavy quark-antiquark 
system at NNLO order:
\begin{eqnarray}
V(r) &=& - \frac { C_F \alpha_s }{r}\Big( 1+\frac {C_F \alpha_s}{4\pi}
\Big( 2\beta_0 \ln(\mu_1 r) + a_1\Big) \\
\nonumber
&+& \Big( \frac{\alpha_s }{4\pi} \Big)^2
\Big( \beta _0^2 \big(  4\ln^2(\mu_1 r)+\frac{\pi^2}{3} \big)
 + 2 (\beta_1 +2\beta_0 a_1) \ln(\mu_1 r) + a_2\Big) \Big) ,
\end{eqnarray}
here $\mu_1=\mu_s e^{\gamma}$, $\mu_s$ is the normalization scale, 
 $\gamma$ is the Euler constant and $\alpha_s=\alpha_s(\mu_s)$.

The coefficients $a_1$ and $a_2$ were calculated in \cite {Fish,Peter},
respectively
\begin{eqnarray}
 a_1 &=& \frac {31}{9} C_A - \frac {20}{9} T_R N_L, \nonumber \\
a_2 &=& \left (\frac {4343}{162} + 6\pi^2 - \frac {\pi^4}{4}+\frac
{22}{3} \zeta_3 \right )C_A^2-\left (\frac {1798}{81}
+\frac {56}{3}\zeta_3\right ) C_A T_R N_L \\ \nonumber
&+& \left (\frac{20}{9} T_R N_L \right )^2
-\left (\frac {55}{3} - 16\zeta_3 \right )C_FT_RN_L.
\end{eqnarray}
The first two coefficients in the expansion of the QCD $\beta$-function
are
\bq
\beta _0 = \frac {11}{3}C_A-\frac {4}{3}N_LT_R,\quad
\beta _1 = \frac {34}{3} C_A^2 - \frac {20}{3} C_AT_RN_L - 4C_F T_R N_L.
\eq
The color factors are
$C_F = 4/3, C_A = 3$ and $T_R = 1/2$. $N_L = 5$ is the number of 
light quarks.\\

{\bf 2b. Breit-Fermi Hamiltonian.}\\

The function $W(r)$ in eq.(\ref{www}) is the QCD generalization of the 
QED Breit-Fermi Hamiltonian \cite{Landau,GRR}. 
We consider here quark-antiquark production in the $S-$wave mode.
The Breit-Fermi Hamiltonian for the final state with $\vec L=0,\vec S^2=2$ 
reads \cite{Landau, GRR}
\be\label{HBF1}
W(r)=-\frac{\vec p^4}{4m^3}+
\frac{11\pi C_F\alpha_s}{3m^2}\delta(\vec r)
-\frac{C_F\alpha_s}{2m^2}\{\frac{1}{r},\vec p^2 \}-
\frac{C_AC_F\alpha_s^2}{2mr^2}.
\ee
Let us demonstrate how the Breit-Fermi Hamiltonian can be reduced to a 
convenient form for the numerical and analytical evaluation.
First we note that eq. (\ref{HBF1}) 
can be rewritten in the following way
\begin{eqnarray}\label{BFH2}
W(r) =-\frac{H_0^2}{4m}+\frac{3}{4m}\{V_0(r), H_0 \}
+\frac{11\pi C_F\alpha_s}{3m^2}\delta(\vec r)
-(\frac{5}{2}+\frac{C_A}{C_F})\frac{V_0^2(r)}{2m}.
\end{eqnarray}
Here we used  $V_0(r)=-\frac{C_F\alpha_s}{r}$. 
The first and second terms in eq.(\ref{BFH2}) can be 
simplified by using equation of motion, 
$ (H-\bar E)G(r',r''|\bar E)=\delta(r'-r''),\quad \bar E=E+i\Gamma$,
but the third and 
fourth terms are related by the commutation relation
\be\label{com}
[H_0,ip_r]=\frac{4\pi\delta(\vec r)}{m}-\frac{C_F\alpha_s}{r^2},
\ee
with  $ip_r=\frac{1}{r}\frac{\partial}{\partial r}r$.
Expressing the term with $1/r^2$ through  
$\delta(\vec r)$ function and using the commutation relations,
 we have (assuming $r\to 0 $)
\bq \label{an1}
G^{NNLO}(r,r|\bar E)&=& (1+\frac{\bar E}{2m})
G^{NNLO}_C(r,r|\bar E+\frac{\bar E^2}{4m})
|_{\alpha_S\to\alpha_s(1+\frac{3\bar E}{2m})}\\ \nonumber
&+&
\frac{4\pi C_F\alpha_s}{3m^2}(1+\frac{3C_A}{2C_F})\Big[ 
G_C^{LO}(r,r|\bar E) \Big]^2.
\eq 
We have omitted some unimportant surface terms above.
Here $G_C(r,r|E)$ is the Green function calculated 
only with the Coulomb 
potential and without the Breit-Fermi Hamiltonian.
 The $LO$ Green function can be written in the form 
\cite{Hoang} 
($v=\sqrt{\frac{\bar E }{ m}}$)
\be 
 G_C^{LO}(0,0|\bar E)=\frac{m^2}{4\pi}\Big(
iv-C_F\alpha_s(\log (\frac{-imv}{\mu_f})+\gamma_E+
\Psi(1-i\frac{C_F\alpha_s}{2v}))\Big),
\ee
where $\mu_f$ is a factorization scale, which disappears in $R$ after 
performing a matching procedure. Employing this form of the LO Green function, 
we get  
the NNLO Green function in the form
\bq \label{an4}
G^{NNLO}(0,0|\bar E)&=&G^{NNLO}_C(0,0|\bar E)-
G^{LO}_C(0,0|\bar E)\\ \nonumber
&+&\frac{4\pi C_F\alpha_s}{3m^2}(1+\frac{3C_A}{2C_F})\Big[ 
G_C^{LO}(0,0|\bar E) \Big]^2  
+\frac{m^2}{4\pi}\Big( iv(1+\frac{5}{8}v^2) \\ \nonumber
&-&\alpha_sC_F(1+2v^2)\Big(\log (\frac{-imv}{\mu_f})+\gamma_E+
\Psi(1-i\frac{C_F\alpha_s}{2v}(1+\frac{11}{8}v^2)) \Big)\Big). \nonumber
\eq
Equations (\ref{an1}) and (\ref{an4}) are in agreement 
with \cite{HT} but they are obtained in a different way.  
Expanding eq.(\ref{an1}) 
in $\alpha_s^2, v\alpha_s, v^2$ we have
\bq \label{an2}
G^{NNLO}(0,0|\bar E)&=&G^{NNLO}_C(0,0|\bar E)+
\delta (G)_1+ \delta (G)_2\\ \nonumber
\delta (G)_1&=&\frac{im^2v}{4\pi}\Big( 
\frac{5}{8}v^2+\frac{11}{16}(\alpha_sC_F)^2\Psi'(1-
i\frac{C_F\alpha_s}{2v})\\ \nonumber
&+&iv2C_F\alpha_s(\log (\frac{-imv}{\mu_f})+\gamma_E+\Psi(1-
i\frac{C_F\alpha_s}{2v}))\Big),\\ \nonumber 
\delta (G)_2&=&\frac{4\pi C_F\alpha_s}{3m^2}(1+\frac{3C_A}{2C_F})\Big[ 
G_C^{LO}(0,0|\bar E) \Big]^2. \nonumber
\eq
These expressions are our analytical results 
for the correction to the Green function at the origin.

An alternative strategy, more consistent one (adopted also in \cite{MY})
is to keep the Breit-Fermi 
correction together with the QCD potential in Schr\"odinger equation
and to solve 
the Schr\"odinger equation numerically. 
The main reason for doing this is because  
the expansion procedure is not safe for energy denominators 
of the Green function
\be
G(r,r|E)=\sum_{n} \frac{|\psi_n(r)|^2 }{E_n-E-i\Gamma}+\int \frac{dk}{2\pi}
\frac{|\psi_k(r)|^2 }{E_k-E-i\Gamma}
\ee
especially if the energy is close to 
the energy of resonant state. 

Therefore, we re-express the $\delta(\vec r)$ function in 
eq.(\ref{BFH2}) by the
term with $1/r^2$ by using the commutation relation (\ref{com})
\bq
W(r) =-\frac{H_0^2}{4m}+\frac{3}{4m}\{V_0(r), H_0 \}
+\frac{11\alpha_sC_F}{12m}[H_0,ip_r] -
(\frac{2}{3}+\frac{C_A}{C_F})\frac{V_0^2(r)}{2m}.
\eq
We consider the first correction to the Green function $G(r',r''|\bar E)$ 
originated from $W(r)$. 
Using the equation of motion we obtain
\bq\label{sur}
-\int d\vec r G(r',r|\bar E)W(r)G(r,r''|\bar E)=
(\frac{\vec p ^2}{2m^2}
+\frac{\alpha_sC_F}{mr'}+ip_r\frac{11\alpha_sC_F}{12 m})
G(r',r''|\bar E) 
+\\ \nonumber
\int d\vec r G(r',r|\bar E)[\frac{\bar E^2}{4m}-\frac{3V_0\bar E}{2m}+
(\frac{2}{3}+\frac{C_A}{C_F})\frac{V_0^2(r)}{2m}]G(r,r''|\bar E).
\eq
The surface term with $ip_r$ does not coincide with the 
one from \cite{MY}.
We see from the second line of the eq.(\ref{sur}) 
that the problem of evaluating the 
Breit-Fermi correction 
is reduced to solving the following equation
\be \label{se1}
(H_1-E_1)G_1(r,r'|E_1)=\delta^3(r-r'),
\ee
with new energy $E_1=\bar E+\frac{\bar E^2}{4m}$ 
and with new Hamiltonian 
\be\label{h1}
H_1=\frac{\vec p^2}{m}+V(r)+\frac{3V_0\bar E}{2m}-
(\frac{2}{3}+\frac{C_A}{C_F})\frac{V_0^2(r)}{2m}.
\ee
The numerical solution of (\ref{se1}) gives us the Green function 
at NNLO order.\\

{\bf 3. Matching procedure. }\\

 Now we consider the matching procedure in order to fix
the short distance coefficient $C(r)$ in eq.(\ref{cross}).
First we calculate the first-order correction to the Green 
function originated from $W(r)$. 
This is given by the following integral
\be
-\int d^3 r G_0(r_0,r|\bar E)W(r)G_0(r,r_0|\bar E)
\ee
with the free Green function (with $k=mv$)
\be
G_0(r,r'|\bar E)=
\frac{m}{4\pi r r'}\Big(\frac{sin(kr)}{k}e^{ikr'}\theta(r'-r)+
\frac{sin(kr')}{k}e^{ikr}\theta(r-r') \Big).
\ee
 Then we compare the result of this integration 
with the NNLO QCD radiative correction to the cross section
$e^+e^-\to t \bar t$ in the region $\alpha_s \ll v \ll 1$  \cite{CM}.
Iterations of the Coulomb potential give only 
terms proportional to $1/v$ or $1/v^2$ at NNLO in this kinematical region.
These terms are trivially identified in $R$.
We obtain the following short distance correction 
\be\label{sdc}
C(r)=1-\frac{4C_F\alpha_s(\mu_h)}{\pi}+
C_2(r)\Big(\frac{C_F\alpha_s(\mu_h)}{\pi}\Big)^2,
\ee
\be
C_2(r)= A_1\log(r/a)+A_2\log(m/\mu_{h})+A_3
\ee
with
\be
A_1=\pi^2(C_A+\frac{2C_F}{3}),
\quad a=\frac{e^{2-\gamma_E}}{2m},\quad A_2=2\beta_0, 
\ee
\begin{eqnarray}
A_3 = C_F C_2^{A} + C_A C_2^{NA}
+T_R N_L C_2^{L} + T_R N_H C_2^{H}
\end{eqnarray}
and with
\begin{eqnarray}\label{sdc2}
C_2^{A}&=&\frac {39}{4} -\zeta _3 +\pi ^2 \left (\frac
{4}{3}\ln2-\frac {35}{18} \right ),
\nonumber \\
C_2^{NA}&=& -\frac {151}{36}-\frac{13}{2} \zeta _3
+\pi^2 \left (\frac {179}{72} -\frac {8}{3}\ln2 \right ),
\nonumber \\
C_2^{H} &=&\frac {44}{9} - \frac {4}{9}\pi^2,
\nonumber \\
C_2^{L} &=& \frac {11}{9}.
\label{hcorr}
\end{eqnarray}
Here $\mu_h$ is the hard normalization scale.
We should note that matching of the analytical expressions 
of the eqs.(\ref{an1})-(\ref{an2}) 
gives similar results but with substitution $r/a \to m/\mu_f$.
These coefficients are in agreement with results obtained in 
\cite{Hoang,HT,MY}. \\

{\bf 4. Numerical results.}\\

 The final expression for the NNLO 
cross section is given by the following equation
\be\label{final}
R_{NNLO}(E)=\Big( 1-\frac{4C_F\alpha_s}{\pi}+
C_2(r_0)\Big(\frac{C_F\alpha_s}{\pi}\Big)^2\Big)
\frac{8\pi}{m^2}{\rm Im}
\Big( (1-\frac{5\bar E}{6m}) G_1(r_0,r_0|E_1) \Big),
\ee
with $G_1(r_0,r_0|E_1)$ being 
solution of the eqs.(\ref{se1})-(\ref{h1}) and
with $C_2(r_0)$ from (\ref{sdc})-(\ref{sdc2}).
The analytical result for $R_{NNLO}$ is given by the same 
formula (\ref{final}) with 
$G(0,0|\bar E)$ from eqs.(\ref{an1})-(\ref{an2})
and with substitutions: $r\to ma/\mu_f$ in $C(r)$ and $(1-
\frac{5\bar E}{6m})\to (1-\frac{4\bar E}{3m})$ in eq.(\ref{final}).
 
In order to solve the Shr\"odinger 
equation numerically we have written two programs
in FORTRAN and MATHEMATICA by
using numerical schemes described in \cite{PS} and \cite{MY}. 
We have checked that the programs reproduce 
the results of the analytical expression for the Green function 
in the pure Coulomb problem and numerical values for NLO cross section 
from \cite{PS,JKTS,HT,MY}. 
The results of both programs for the 
NNLO correction
give the same result; therefore we are confident in our numerical results.

The Fig.1 shows our final results for $R_{NNLO}(E)$ as a function of
the nonrelativistic energy  
$E=\sqrt{s}-2m$. We compare the NNLO results with LO and NLO curves, 
for the soft scale $\mu_s=50,75,100$ GeV.
We have chosen $m=175 GeV$, 
$\Gamma =1.43 GeV$, $r_0=a$ and $\alpha_s(M_Z)=0.118$.
We see that NNLO correction is  
of the order $20\%$ and as large as NLO one.
The $1S$ peak is shifted towards smaller energies.
The dependence of the NNLO cross section on the parameter $\mu_s$ is 
weaker than in LO, but more robust than in NLO. 

The dependences on the factorization scale $r_0$ and hard 
normalization scale  $\mu_{h}$
are much smaller than the dependence on $\mu_s$. 
We demonstrate that fact in Fig 2., where
we plot $R_{NNLO}(E)$ as a function of energy  
at $r_0=2a , a , a/2$. In Fig.3  we show  $R_{NNLO}(E)$ 
at $\mu_{h}=2m, m, m/2 $. 
In Fig.4  we demonstrate $R_{NNLO}(E)$ at different 
$\alpha_s(M_Z)=0.116, 0.118, 0.12$.
We see that $R_{NNLO}(E)$ is very sensitive to the value of the 
QCD coupling $\alpha_s(M_Z)$.
Comparing our results (Fig.1) with the numerical 
results of \cite{MY} 
we have found that they differ by about $4-6\%$ at the 
$1S$ peak and above it 
and by $10-20\%$ bellow the $1S$ peak, where the cross 
section is quit small and NNLO correction is large. 
The difference of our results and the results of \cite{MY} 
appears because of error in numerical solution of the Schr\"odinger equation. 

If we expand the Green function in the 
Breit-Fermi Hamiltonian we obtain the result that is in agreement 
with result presented in \cite{HT}, see eqs. 
(\ref{an1})-(\ref{an2}). 
However, the expansion of the energy denominator 
of the Green function is not correct in the resonance energy region,
where an expansion parameter, $\frac{\delta E_n}{E-E_n}$, is large.
We do not use it in our calculation, 
and prefer to solve Schr\"odinger equation numerically
with effective potential contained the Coulomb potential and
the Breit-Fermi potential.
Therefore, our final result differs numerically from both 
results presented\footnote{
After this paper has been accepted for publication,
the authors of 
\cite{MY} have agreed with the 
numerical results obtained in this paper.

} in \cite{HT} and \cite{MY}.
The qualitative conclusion about the large size and the positive sign 
of the NNLO correction is in agreement with \cite{HT} and \cite{MY}.

{\bf 5. Conclusion and outlook.} \\
In conclusion, we have calculated the total cross section,
resumming Coulomb-like terms with next-to-next-to-leading   
accuracy. We  used the nonrelativistic Green function
formalism and the method of ``direct matching''.
The NNLO correction turns out to be large, of the same 
order as the NLO correction. It shifts the position of the $1S-$peak 
and changes the its normalization.

The comparison of obtained results with the results published in \cite{HT,MY} 
shows that our final results are in qualitative agreement, but 
differ numerically from that of \cite{HT} and \cite{MY}.

Finally, let us mention some open questions in this field.
It is quite important to consider NNLO correction to the 
total cross
section in the scheme with the running QCD coupling $\alpha_s$ and
in the scheme with the low-scale running top quark mass, 
for example $m_{PS}(\mu)$ \cite{Beneke1}.                     
It would be interesting in future to examine the size 
of the NNLO corrections to the differential distributions,
for example to the distribution over spatial momentum, 
$\frac{d\sigma}{d|\vec p_{t\bar t}|}$.
It is necessary to analyze nonfactorizable corrections to 
the differential cross section at NNLO. 
In view of the fact that NNLO correction is large it would 
be useful to check NNLO correction to the static QCD potential, 
the coefficient $a_2$ \cite{Peter}.\\



{\bf Acknowledgements}.

I am grateful to A. Khodjamirian, R. R\"uckl
for discussions and comments.
This work is supported by the German Federal Ministry for 
Research and Technology (BMBF) under contract number 05 7WZ91P (0).

\begin{figure}
\centerline{
\epsfig{
file=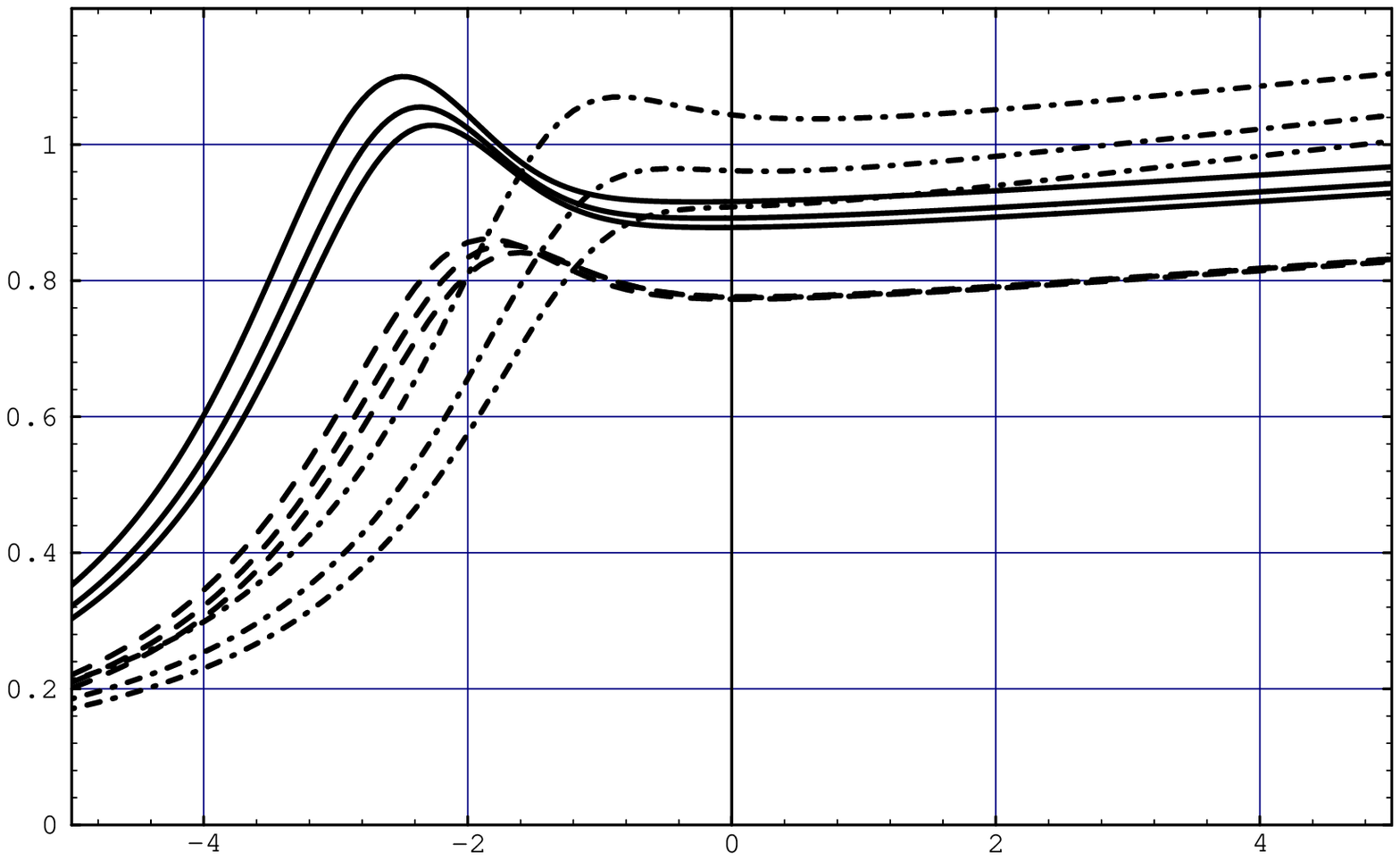,
scale=0.7,%
clip=}}
\caption{$R(e^+e^-\to t\bar t)$ for the 
LO (dashed-dotted lines ), NLO (dashed lines), NNLO (solid lines) 
approximation  as a function of energy $E=\sqrt{s}-2m$, GeV.
In all cases we use $m_t=$175 GeV,$\Gamma_t=$1.43 GeV, 
$\alpha_s(m_Z)=$0.118 but different
values of the soft scale $\mu_s=50,75,100$~GeV~.}
\end{figure}

\begin{figure}
\centerline{
\epsfig{
file=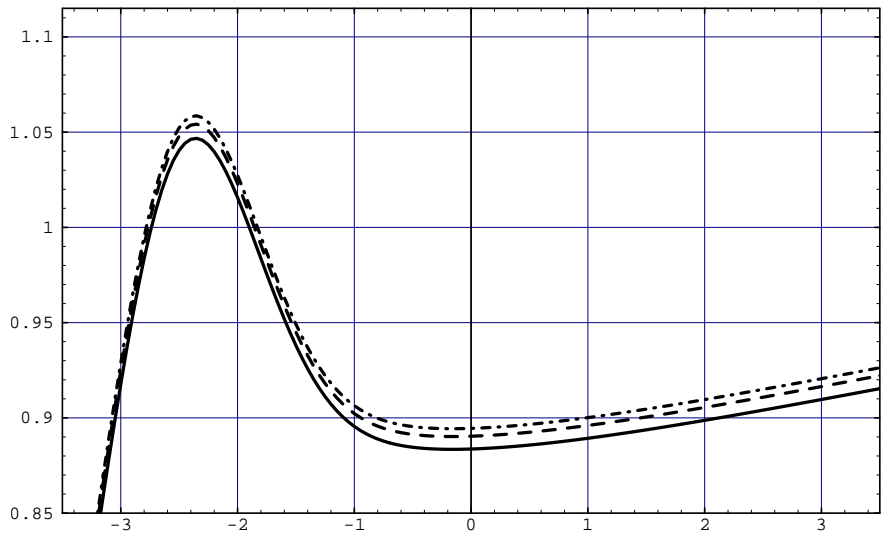,
scale=0.7,%
clip=}}
\caption{$R(e^+e^-\to t\bar t)$ at NNLO  at different values of
factorization scale $r_0=2a$(dashed-dotted line),
$r_0=a$ (dashed line) and $r_0=a/2$ (solid line).
We use $m_t=$175 GeV, $\Gamma_t=$1.43 GeV, 
$\mu_s=75$ GeV,$\alpha_s(m_Z)=$0.118. }
\end{figure}
\begin{figure}
\centerline{
\epsfig{
file=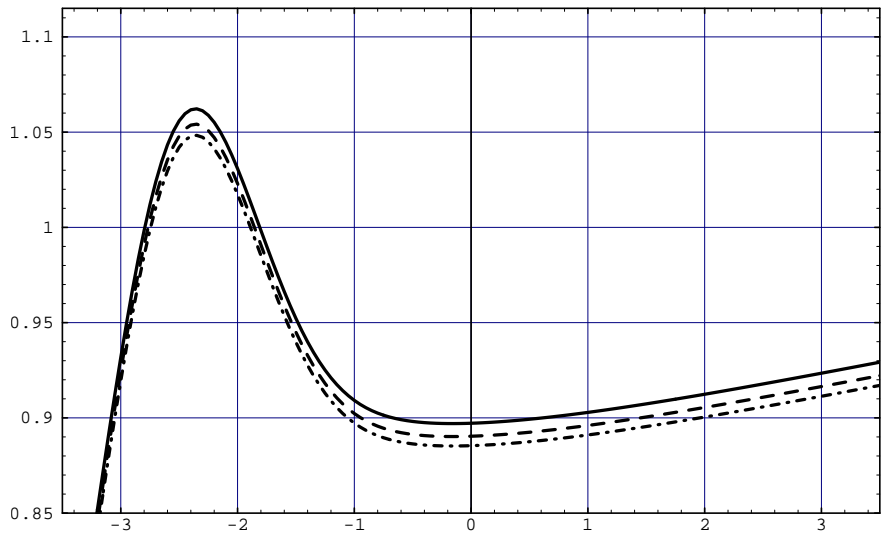,
scale=0.7,%
clip=}}
\caption{$R(e^+e^-\to t\bar t)$ at NNLO  at different value
of the hard normalization scale $\mu_h=2m$(dashed-dotted line),
$\mu_h=m$ (dashed line) and $\mu_h=m/2$ (solid line).
We use $m_t=$175 GeV, $\Gamma_t=$1.43 GeV, 
$\mu_s=75$ GeV,$\alpha_s(m_Z)=$0.118.}
\end{figure}

\begin{figure}
\centerline{
\epsfig{
file=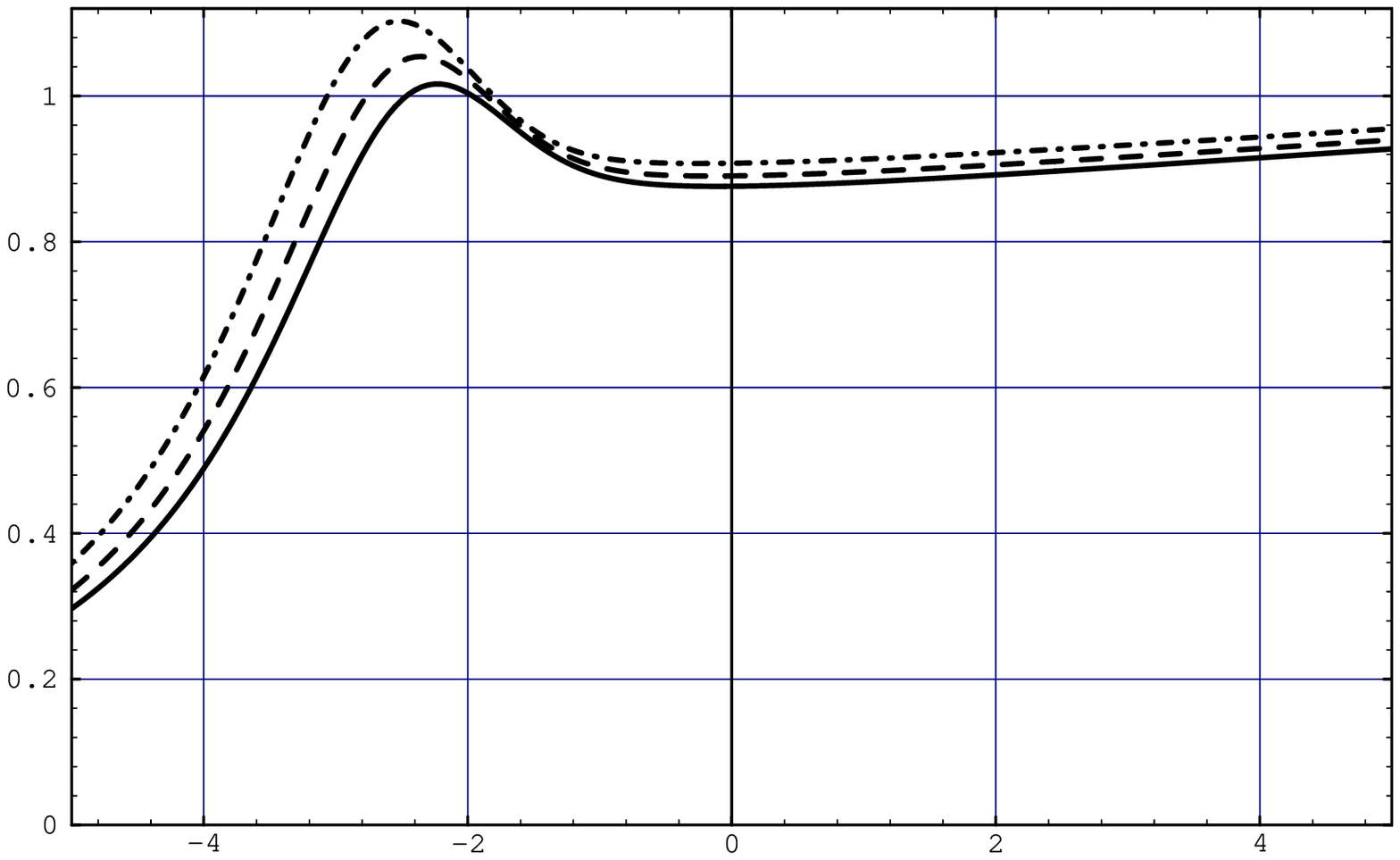,
scale=0.7,%
clip=}}
\caption{$R(e^+e^-\to t\bar t)$ at NNLO at different values of the QCD
coupling constant $\alpha_s(m_Z)=$0.116 (solid line),
$\alpha_s(m_Z)=$0.118 (dashed line),$\alpha_s(m_Z)=$0.120    
(dashed-dotted line).
We use $m_t=$175 GeV, $\Gamma_t=$1.43 GeV and $\mu_s=75$ GeV.}
\end{figure}

\newpage



\begin{thebibliography}{99}
\bibitem{Rev} E. Accomando {\it et al.}, the ECFA/DESY LC Physics
 Working Group, \\
{\em Physics with $e^+ e^-$ linear colliders}, DESY 
97-100, hep-ph/9705442.
\bibitem{FK} V.S. Fadin and V.A. Khoze,  JETP Lett. {\bf 46}
(1987), 525; \\
Sov. J. Nucl. Phys. {\bf 48} (1988), 309.
\bibitem{PS} M. Peskin and M. Strassler, 
Phys. Rev. {\bf D43} (1991), 1500.
\bibitem{WK} W. Kwong, Phys. Rev.{\bf D43} (1991), 1488.
\bibitem{JKTS} M.Je\'zabek, J.H.K\"uhn and
T. Teubner, Z. Physik {\bf C56} (1992), 653;\\
Y. Sumino, K. Fujii, K. Hagiwara, H. Murayama and C.-K. Ng,\\
Phys. Rev. {\bf D47} (1993), 56.
\bibitem{Sumino} Y. Sumino, Ph.D thesis, Tokyo, 1993;
\bibitem{MYa} K. Melnikov and O. Yakovlev, 
Phys. Lett. {\bf B324} (1994), 217.
\bibitem{Hoang} A.H. Hoang, Phys. Rev. {\bf D56} (1997), 5851.
\bibitem{Kuehn} A.H. Hoang, J. H. K\"uhn and T. Teubner, 
Nucl. Phys. {\bf B 452} (1995) 173.
\bibitem{CM} A. Czarnecki and 
K.Melnikov, Phys. Rev. Lett. {\bf 80} (1998) 2531.
\bibitem{Beneke} M. Beneke and V. A. Smirnov, 
Nucl. Phys. {\bf B522} (1998) 321\\
M. Beneke, A.Signer and V.A. Smirnov, Phys. Rev. Lett. 80
(1998) 2535.
\bibitem{HT} A.H. Hoang and T. Teubner, preprint UCSD/PTH 98-01,\\
DESY 98-008, hep-ph/9801397.
\bibitem{MY} K. Melnikov and A. Yelkhovsky,
Nucl.Phys. {\bf B528} (1998) 59. 
\bibitem{Fadin} V.S Fadin, 
V.A. Khoze, A.D. Martin, Phys. Rev. {\bf D49} (1994), 2247.
\bibitem{FKM1}  V.S Fadin, 
V.A. Khoze, A.D. Martin,  Phys. Lett. {\bf B320} (1994), 141. 
\bibitem{Fish} W. Fischler, Nucl. Phys. {\bf B129} (1977), 157;\\
A. Billoire, Phys. Lett. {\bf B92} (1980), 343.
\bibitem{Peter} M. Peter, Nucl. Phys. {\bf B501} (1997), 471.
\bibitem{Landau} L.D. Landau and  E.M. Lifschitz,
 {\it Relativistic Quantum Theory}, \\
part 1 (Pergamon, Oxford, 1974).
\bibitem{GRR} S.N. Gupta and S.F.Randford, Phys. Rev. {\bf D24}
(1981), 2309,\\ (E) {\it ibid} {\bf D25} (1982), 3430;\\
S.N. Gupta, S.F.Randford and W.W. Repko, Phys. Rev. {\bf D26} (1982), 3305.
\bibitem{Beneke1} M. Beneke, preprint CERN-TH/98-120, hep-ph/9804241.  
\end{thebibliography}
\end{document}